\documentclass[prb,aps,tightenlines,twocolumn,
floatfix]{revtex4}
\usepackage{graphicx}
\usepackage{amssymb}
\usepackage{amsmath}
\usepackage{bm}
\usepackage{colordvi}
\usepackage{mathrsfs}
\usepackage{cases}
\def\up{\uparrow}
\def\down{\downarrow}

\begin{document}
\title{Comment on ``Magnetotransport through graphene spin valves'' and its
  following works} 
\author{Y. Zhou}
\author{M. W. Wu}
\thanks{Author to whom correspondence should be addressed}
\email{mwwu@ustc.edu.cn.}
\affiliation{Hefei National Laboratory for Physical Sciences at
  Microscale and Department of Physics, University of Science and
  Technology of China, Hefei, Anhui, 230026, China}

\date{\today}
\begin{abstract}
  We show that the key formula in the works 
  [Ding, Zhu, and Berakdar, Phys. Rev. B {\bf 79}, 045405 (2009); 
  {\bf 84}, 115433 (2011); Ding, Zhu, Zhang, and Berakdar,
  Phys. Rev. B {\bf 82}, 155143 (2010)]
  is invalid in the extended graphene system they investigated.
  The correct formalism in the extended infinite system is also
  presented in this comment. 
\end{abstract}

\maketitle

Recently, Ding {\em et al.} presented a series of investigations on
the spin-polarized transport in graphene contacted to ferromagnetic
electrodes.\cite{Berakdar_09,Berakdar_10,Berakdar_11} 
Different from the usual works on the mesoscopic transport,
they claimed that the system in their model is an extended graphene layer, i.e.,
an infinite system.\cite{Berakdar_comment} 
However, their key formula [e.g., Eq.~(11) in Ref.~\onlinecite{Berakdar_09}] is
only valid in the finite system but fails in their system. 
In the following, we address this problem in detail
and further give the correct formalism in the extended infinite system.

We first show that the formulae by Ding {\em et
  al.}\cite{Berakdar_09,Berakdar_10,Berakdar_11}  
are invalid in the infinite system.
Here we only discuss the spinless case without time-dependent
fields,\cite{Berakdar_09} but our main conclusion also applies to the more 
complicated cases, e.g., the cases with ferromagnetic electrodes and/or under
the time-dependent fields discussed in
Refs.~\onlinecite{Berakdar_09}-\onlinecite{Berakdar_11}.  
The main formula used in their calculations [Eq.~(11) in
Ref.~\onlinecite{Berakdar_09}] reads 
\begin{equation}
  I_L=-\frac{e}{h}\int {d\varepsilon}\;
  {\cal G}_{a}^{r}(\varepsilon) \Gamma_{R}(\varepsilon)
  {\cal G}_{a}^{a}(\varepsilon) \Gamma_{L}(\varepsilon)
  [f_L(\varepsilon)-f_R(\varepsilon)],
  \label{I_wrong}
\end{equation}
where ${\cal G}_{a}^{r,a,>,<}(\varepsilon)=\frac{1}{N}\sum\limits_{\mathbf{qq'}}
G_{\mathbf{q}a,\mathbf{q'}a}^{r,a,>,<}(\varepsilon)$ with 
$G_{\mathbf{q}a,\mathbf{q'}a}^{r,a,>,<}(\varepsilon)$ being the retarded, 
advanced, greater and lesser Green's functions of graphene connecting with leads
and $N$ being the number of sites on A sublattice. 
Note that they assumed that only atoms in the A sublattice are connected with
the electrodes, and the momentum dependence of the self-energy can be
neglected. In order to obtain the above formula, they have to use
two relations,\cite{Berakdar_09} 
\begin{numcases} {}
  {\cal G}_a^r(\varepsilon)-{\cal G}_a^a(\varepsilon)
  ={\cal G}_a^r(\varepsilon)
  [{\Sigma}^r(\varepsilon)-{\Sigma}^a(\varepsilon)]
  {\cal G}_a^a(\varepsilon), 
  \label{G_r_rel2} 
  \\ [3pt]
  {\cal G}_a^{>,<}(\varepsilon)={\cal G}_a^r(\varepsilon)
  {\Sigma}^{>,<}(\varepsilon) {\cal G}_a^a(\varepsilon).
  \label{G_less_rel2}
\end{numcases} 
These relations are indeed valid in the finite
system.\cite{Datta_mesoscopic,provement} However, they are invalid in the 
infinite system as shown in the following.

From the Dyson equation
\begin{equation}
  {\cal G}_a^{r,a}(\varepsilon)=\overline{g}_a^{r,a}(\varepsilon)
  + \overline{g}_a^{r,a}(\varepsilon) {\Sigma}^{r,a}(\varepsilon)
  {\cal G}_a^{r,a}(\varepsilon).
  \label{Dyson}
\end{equation}
where $\overline{g}_{a}^{r,a}(\varepsilon)
=\frac{1}{N}\sum\limits_\mathbf{q}g_{\mathbf{q}a,\mathbf{q}a}^{r,a}(\varepsilon)$
with $g_{\mathbf{q}a,\mathbf{q'}a}^{r,a,>,<}(\varepsilon)$ being the retarded,
advanced, greater and lesser Green's functions of graphene without connecting
with leads, one obtains,
\begin{eqnarray}
  {\cal G}_a^r(\varepsilon)-{\cal G}_a^a(\varepsilon)&=&
  {\cal G}_a^r(\varepsilon)[{\Sigma}^r(\varepsilon)
  -{\Sigma}^a(\varepsilon)]{\cal G}_a^a(\varepsilon)
  \nonumber \\ &&
  {} - {\cal G}_a^r(\varepsilon)[\overline{g}_a^r(\varepsilon)^{-1}
  -\overline{g}_a^a(\varepsilon)^{-1}]{\cal G}_a^a(\varepsilon).\quad
  \label{G_r_correct}
\end{eqnarray}
In the finite system,
\begin{eqnarray}
  && \overline{g}_a^{r,a}(\varepsilon)=\frac{1}{N}\frac{B_1}{B_0\pm B_1i0^+},\\
  && \hspace{-0.8cm} 
  B_0=\prod\limits_{i}(\varepsilon-\epsilon_{i}), \qquad
  B_1=\sum_{j} \prod\limits_{i\ne j} (\varepsilon-\epsilon_{i}),
\end{eqnarray}
where $i$ represents the joint index of the momentum index ${\bf q}$ and the
band index $\eta$. 
Thus, 
\begin{equation}
  \overline{g}_a^r(\varepsilon)^{-1}-\overline{g}_a^a(\varepsilon)^{-1}
  = 2N i0^+ = 0.
\end{equation}
This indicates that the second term in
Eq.~(\ref{G_r_correct}) vanishes and hence Eq.~(\ref{G_r_rel2}) is valid in the
finite system, in consistence with the previous
literature.\cite{Datta_mesoscopic,provement} However, in the infinite 
system discussed by Ding {\em et al.},\cite{Berakdar_09,Berakdar_10,Berakdar_11} 
the situation becomes totally different.
After converting the summation over ${\bf q}$ in $\overline{g}_a^{r,a}(\varepsilon)$
into the integration in two-dimensional momentum space, one obtains
\begin{eqnarray}
  \overline{g}_{a}^{r,a}(\varepsilon)
  &=&-F_0(\varepsilon)\mp \frac{i}{2} A_0(\varepsilon), 
  \label{g_small} \\
  F_0(\varepsilon)&=&
  \frac{S_0\varepsilon}{2\pi\hbar^2 v_{\rm F}^2}\ln\frac{|\varepsilon^2-D^2|}{\varepsilon^2}, 
  \label{F_0} \\
  A_0(\varepsilon)&=&\frac{S_0|\varepsilon|}{\hbar^2v_{\rm F}^2}\theta(D-|\varepsilon|),
  \label{rho_0}
\end{eqnarray}
where $D$ is the cutoff energy and $S_0$ is the area of the unit cell. 
The above formulae give 
$\overline{g}_a^r(\varepsilon)^{-1}-\overline{g}_a^a(\varepsilon)^{-1}$ 
is nonzero for $0<|\varepsilon|<D$. Therefore, Eq.~(\ref{G_r_rel2}) is
invalid and should be replaced by Eq.~(\ref{G_r_correct}). 
Further considering that 
\begin{equation}
  {\cal G}^>_a(\varepsilon)-{\cal G}^<_a(\varepsilon)
  ={\cal G}^r_a(\varepsilon)-{\cal G}^a_a(\varepsilon),
  \label{G_less_G_r}
\end{equation}
which is directly from the definition of the Green's functions, one can
conclude that Eq.~(\ref{G_less_rel2}) is also invalid in their system. 
Therefore, their main formula Eq.~(\ref{I_wrong}) is incorrect, and all their
following results lose the scientific ground.

\begin{figure}[tbp]
  \begin{center}
    \includegraphics[width=7.cm]{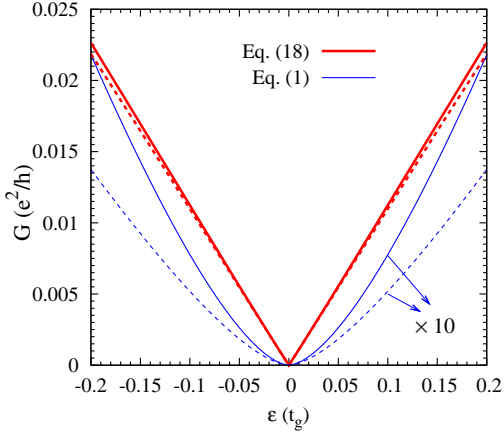}
  \end{center}
  \caption{ (Color online) The differential conductances from our formula
    [Eq.~(\ref{I_final}), red curves] and those from the  
    formula by Ding {\em et al.}\cite{Berakdar_09,Berakdar_10,Berakdar_11}
    [Eq.~(\ref{I_wrong}), blue curves] as function of bias
    for $D=2.3 t_g$ (the Debye cutoff, solid curves) and $t_g$ (dashed curves),
    $\Gamma_{L}=\Gamma_{R}=\Gamma_0$ with $\Gamma_0=0.5t_g$. 
    In order to better comparing the results from these two formulae,
    we even have to multiply the results from Eq.~(\ref{I_wrong}) by a
    factor 10. 
  }
  \label{fig_conduct} 
\end{figure}

Then we further give the correct formulae in the extended infinite system. 
Our starting point is Eq.~(8) in Ref.~\onlinecite{Berakdar_09}, which 
can be written as
\begin{equation}
  I_\alpha=-\frac{ie}{h}\int{d\varepsilon}\;
  \Big\{ [{\cal G}_{a}^{r}(\varepsilon)
  -{\cal G}_{a}^{a}(\varepsilon)]f_\alpha(\varepsilon)
  + {\cal G}_{a}^{<}(\varepsilon) \Big\} \Gamma_{\alpha}(\varepsilon),
  \label{I_L} 
\end{equation}
where $\alpha=L,R$ stand for the left and right leads; the chemical potential in
the leads are $\mu_{L,R}=\pm eV/2$. 
From the steady-state condition $I_L+I_R=0$ and Eq.~(\ref{G_less_G_r}), one
obtains the correct formula of ${\cal G}^{>,<}_a(\varepsilon)$ for the infinite
system, 
\begin{equation}
  {\cal G}_{a}^{>,<}(\varepsilon)=\frac{i{\Sigma}^{>,<}(\varepsilon)}
  {\Gamma_{L}(\varepsilon)+\Gamma_{R}(\varepsilon)}
  [{\cal G}_{a}^{r}(\varepsilon)-{\cal G}_{a}^{a}(\varepsilon)].
  \label{G_less_correct}
\end{equation}
In order to compare the above formula with the incorrect one
[Eq.~(\ref{G_less_rel2})], we rewrite Eq.~(\ref{G_less_correct})
into the following form
\begin{eqnarray}
  {\cal G}_{a}^{>,<}(\varepsilon) &=& [1+{\cal G}_a^r(\varepsilon) 
  {\Sigma}^r(\varepsilon)]\overline{g}_a^{>,<}(\varepsilon)
  [1+{\Sigma}^a(\varepsilon){\cal G}_a^a(\varepsilon)] 
  \nonumber \\ && 
  {} +{\cal G}_a^r(\varepsilon) {\Sigma}^{>,<}(\varepsilon) 
  {\cal G}_a^a(\varepsilon),\\
  \label{G_less_full}
  \overline{g}_a^{>,<}(\varepsilon)&=&
  \mp i A_0(\varepsilon) f_g^{>,<}(\varepsilon),\\
  f_g^{>,<}(\varepsilon)&=&\frac{\Gamma_{L}(\varepsilon)f_L^{>,<}(\varepsilon)
    +\Gamma_{R}(\varepsilon)f_R^{>,<}(\varepsilon)}
  {\Gamma_{L}(\varepsilon)+\Gamma_{R}(\varepsilon)},
\end{eqnarray}
with $f_\alpha^<(\varepsilon)=f_\alpha(\varepsilon)$ and 
$f_\alpha^>(\varepsilon)=1-f_\alpha(\varepsilon)$. 
The comparison of Eqs.~(\ref{G_less_rel2}) and (\ref{G_less_full}) shows
that the term related to the graphene distribution $f_g^{<}(\varepsilon)$ must
be taken into account in the infinite system, although this term vanishes in the
finite system as shown in the literature.\cite{provement} 
This again reflects the distinct properties in the infinite and finite systems.

From Eqs.~(\ref{I_L}) and (\ref{G_less_correct}), one further obtains the
correct formula of the current for extended graphene system
\begin{eqnarray}
  I_L&=&-\frac{ie}{h}\int {d\varepsilon}\;
  \frac{\Gamma_{L}(\varepsilon)\Gamma_{R}(\varepsilon)}
  {\Gamma_{L}(\varepsilon)+\Gamma_{R}(\varepsilon)}
  [{\cal G}_{a}^{r}(\varepsilon)-{\cal G}_{a}^{a}(\varepsilon)]
  \nonumber \\ &&
  {} \times [f_L(\varepsilon)-f_R(\varepsilon)].
  \label{I_final}
\end{eqnarray}
In fact, the difference between the above formula and the one by Ding 
{\em et al.}\cite{Berakdar_09,Berakdar_10,Berakdar_11} [Eq.~(\ref{I_wrong})] is
just from the second term in Eq.~(\ref{G_r_correct}). 
As mentioned above, in the finite system, the second term in
 Eq.~(\ref{G_r_correct}) vanishes, thus
Eq.~(\ref{I_final}) is equivalent to Eq.~(\ref{I_wrong}). 
However, this additional term in  Eq.~(\ref{G_r_correct})
becomes very significant in the infinite system. 
In order to make this issue more pronounced, we plot the differential
conductances $G=-dI_L/dV$ obtained from Eqs.~(\ref{I_final}) and (\ref{I_wrong})
in Fig.~\ref{fig_conduct}. Here we adopt the cutoff following the Debye
prescription:\cite{Berakdar_09,Berakdar_10,Berakdar_11,Neto_PRL,equal} the cutoff
energy is chosen to ensure the conservation of the total number of states in the
Brillouin zone after linearization of the spectrum around the K point,
which gives $D=2.3t_g$. 
The other parameters are set to be $\Gamma_{L}=\Gamma_{R}=\Gamma_0$ with
$\Gamma_0=0.5t_g\approx 0.2 D$, just as those in Ref.~\onlinecite{Berakdar_10}.
From this figure, one observes that the results from our formula
(red solid curve) are over one order of magnitude larger than those from the
formula by Ding  {\em et al.}\cite{Berakdar_09,Berakdar_10,Berakdar_11} 
(blue solid curve) in low-energy regime. 
This clearly justifies the importance of the term missed by Ding 
{\em et al.}\cite{Berakdar_09,Berakdar_10,Berakdar_11} and 
further confirms the incorrectness of Eq.~(\ref{I_wrong}).
Considering all the results in
Refs.~\onlinecite{Berakdar_09}-\onlinecite{Berakdar_11} are based on the above 
incorrect formula, one can conclude that all their results are incorrect. 

Finally, we address the treatment of the cutoff.
The precondition of the Debye cutoff used above and in fact any cutoff
is that the physics in the long-wavelength (low-energy) limit is
irrelevant to the high-energy states. 
This is because under this condition, even after the genuine
dispersion at high energy is considered and the value of the cutoff
energy is replaced, the physics around the Dirac point keeps 
the same.
In order to examine whether the above condition is satisfied in the two
different formalisms, we plot the conductances from Eqs.~(\ref{I_final}) 
and (\ref{I_wrong}) with another value of cutoff energy $D=t_g$ in 
Fig.~\ref{fig_conduct}. 
It is seen that the results from our formula are indeed independent of
the cutoff energy when the energy is far below the cutoff energy, e.g., the
relative difference between the results with two values of $D$ is smaller than
$2.5\%$ for $|\varepsilon|<0.1t_g$.
However, the results from the formula by Ding 
{\em et al.}\cite{Berakdar_09,Berakdar_10,Berakdar_11} are shown to depend on
the cutoff energy even in the energy range far below the cutoff energy, e.g.,
the relative difference is over $20\%$ for $|\varepsilon|=0.01t_g$.
The above comparison indicates that our formalism indeed satisfies the
condition of the cutoff approximation but the approach by Ding {\em et
  al.}\cite{Berakdar_09,Berakdar_10,Berakdar_11} fails.
This is another strong evidence to judge which approach is correct.
Moreover, as pointed out in our previous work,\cite{Zhou_pump,Zhou_comment} 
the most important features in their work,\cite{Berakdar_11} such as the
pronounced peaks in the transmissions, come from the contribution of the 
states around and even beyond the cutoff energy. Obviously, such
treatment is unacceptable for an approach introducing a cutoff.

{\em Note added in proof}. We would like to point out that there is
a serious mistake in the reply of Ding {\em et al.} to this comment. In Fig. 2
in that reply, they show that the 
tunnel magnetoresistance (TMR) from our approach becomes zero at zero
bias. However, it is impossible to obtain such a result through our formula
[Eq.~(18)]. This can be proved as follows. From Eqs.~(9)-(11), one obtains 
$|\Gamma_0\overline{g}^{r,a}_a(\varepsilon)|\ll 1$ for $\varepsilon\approx 0$. 
Thus, ${\cal G}_a^{r,a}(\varepsilon)=\overline{g}^{r,a}_a(\varepsilon)/[1-
{\Sigma}^{r,a}(\varepsilon)\overline{g}_a^{r,a}(\varepsilon)]\approx
\overline{g}^{r,a}_a(\varepsilon)$. Further exploiting Eq.~(18) and the
definition of the TMR $[I(0)-I(\pi)]/I(0)$ with $I(0)$ [$I(\pi)$] being the
current in the parallel (antiparallel) configuration of the electrode
magnetization, one obtains the TMR in the zero-bias limit 
[here
$\gamma_\sigma(\theta)=\Gamma_L^\sigma \Gamma_R^\sigma(\theta) /(\Gamma_L^\sigma
+ \Gamma_R^\sigma(\theta))$]   
${\rm TMR}={\sum_\sigma[\gamma_\sigma(0)-\gamma_\sigma(\pi)]} /
{\sum_\sigma\gamma_\sigma(0)}
={(\Gamma_L^\up-\Gamma_L^\down)^2}/{(\Gamma_L^\up+\Gamma_L^\down)^2}$.
Obviously, this TMR is nonzero for nonzero spin polarization
($\Gamma_L^\up\ne\Gamma_L^\down$) used in the reply by Ding {\em et al.}. 
Apparently, some errors are made in their computation.

This work was supported by the National Basic Research Program of China under
Grant No.\ 2012CB922002 and the Strategic Priority Research Program of the
Chinese Academy of Sciences under Grant No. XDB01000000.


\begin{thebibliography}{0}
\bibitem{Berakdar_09} K.-H. Ding, Z.-G. Zhu, and J. Berakdar, Phys. Rev. B 
  {\bf 79}, 045405 (2009).
\bibitem{Berakdar_10} K.-H. Ding, Z.-G. Zhu, Z.-H. Zhang, and J. Berakdar,
  Phys. Rev. B {\bf 82}, 155143 (2010).
\bibitem{Berakdar_11} K.-H. Ding, Z.-G. Zhu, and J. Berakdar, Phys. Rev. B 
  {\bf 84}, 115433 (2011).
\bibitem{Berakdar_comment} Z.-G. Zhu and J. Berakdar, arXiv:1207.3457v1;
  see also Ref.~\onlinecite{Zhou_comment}.
\bibitem{Zhou_comment} Y. Zhou and M. W. Wu, arXiv:1207.4365v1.
\bibitem{Datta_mesoscopic} S. Datta, {\em Electronic Transport in Mesoscopic 
    Systems} (Cambridge University Press, Cambridge, 1995).
\bibitem{provement} J. Maciejko, ``An Introduction to Non-Equilibrium Many-Body
  Theory'', available for download at
  http://www.physics.arizona.edu/\~{}stafford/Courses/560A/
  nonequilibrium.pdf.
\bibitem{Neto_PRL} B. Uchoa, V. N. Kotov, N. M. R. Peres, and A. H. Castro Neto, 
  Phys. Rev. Lett. {\bf 101}, 026805 (2008).
\bibitem{equal} For the Debye cutoff, $S_0/(2\pi \hbar^2 v_{\rm F}^2)=1/D^2$. In
  this case, Eqs.~(\ref{g_small})-(\ref{rho_0}) are just equivalent 
  to Eqs.~(13) and (14) in Ref.~\onlinecite{Berakdar_09}. 
\bibitem{Zhou_pump} Y. Zhou and M. W. Wu, Phys. Rev. B {\bf 86}, 085406 (2012). 
\end{thebibliography}
\end{document}